\documentclass[prd,preprintnumbers,floatfix,aps,nofootinbib,notitlepage,twocolumn]{revtex4}

\usepackage{latexsym}
\usepackage{epsfig}
\usepackage{amssymb}

\newcommand{\lp}{\left(}
\newcommand{\rp}{\right)}
\newcommand{\lb}{\left[}
\newcommand{\rb}{\right]}

\newcommand{\ba}{\begin{eqnarray}}
\newcommand{\ea}{\end{eqnarray}}
\newcommand{\be}{\begin{equation}}
\newcommand{\ee}{\end{equation}}

\newcommand{\al}{\alpha}

\newcommand{\ga}{\gamma}

\newcommand{\la}{\lambda}

\newcommand{\R}{\mathcal{R}}

\begin{document}

\title{The post-Newtonian limit in C-theories of gravitation}

\author{Tomi S. Koivisto}
\email{tomi.koivisto@fys.uio.no}
\affiliation{Institute for Theoretical Physics and Spinoza Institute, Leuvenlaan 4, 3584 CE Utrecht, The Netherlands.}

\date{\today}

\begin{abstract}

C-theory provides a unified framework to study metric, metric-affine and more general theories of gravity.
In the vacuum weak-field limit of these theories, the parameterized post-Newtonian (PPN) parameters $\beta$ and $\gamma$ can differ
from their general relativistic values. However, there are several classes of models featuring long-distance modifications
of gravity but nevertheless passing the Solar system tests. Here it is shown how compute the PPN parameters in C-theories and also in 
nonminimally coupled curvature theories, correcting previous results in the literature for the latter.

\end{abstract}

\maketitle

\section{Introduction}

There are many fascinating applications of modified gravity to cosmology \cite{Nojiri:2006ri,Clifton:2011jh}. In the context of more general gravity theories than Einstein's General 
Relativity (GR), one may hope to resolve such possibly deep issues in the standard model of cosmology as the initial singularity and the subsequent inflation, 
or the dark sector of the universe that consists of dark matter and the cosmological constant. Most importantly, with the present and forthcoming experimental data, 
we are able to observationally test various aspects of the nature of gravity at cosmological scales to high precision. However, any viable theory of gravity should 
of course be able to reproduce the successes of GR at the near-Newtonian scales relevant at the Solar system, where we already have tight bounds on the deviations 
from GR. 

Here we examine the C-theories of gravitation \cite{Amendola:2010bk} from this local point of view. First in section \ref{limits} we write down the C-theory action and 
review some results concerning special classes of theories it contains. In particular, we point out that the previous derivations of the PPN parameters in the 
nonminimally coupled class of models (\ref{nmc}) are incorrect. In section \ref{params} we present the scalar-tensor formulation of the C-theory action and compute 
the corresponding PPN parameters. We show there are three qualitatively different ways to reconcile these theories with the Solar system experiments 
without resorting to the chameleon mechanism. Some explicit results for specific models are given in the appendix \ref{app}.

\section{On some limits of $C(R)$ theory}
\label{limits}

\subsection{The C-theory action}

Any physical theory of gravitation contains two affine structures. One connection determines the geometry, another the geodesics of matter. In GR these connections coincide. 
The C-theories emerge from a nontrivial relation between the connections \cite{Amendola:2010bk}. In the simplest case, which can be more specifically referred to as the $C(\R)$ 
theories, the relation between the connections is conformal and depends solely upon the curvature scalar $\R$. The action can then be written as 
\be \label{action}
S_C = \int d^4 x\sqrt{-g} \lb f(\R) + \hat{\la} - C(\R){\la} + \mathcal{L}_m \rb\,,
\ee
where $\la = \la^{\mu\nu}g_{\mu\nu}$, $\hat{\la} = \la^{\mu\nu}\hat{g}_{\mu\nu}$ and
\be
\R  \equiv g^{\mu\nu}\lp \hat{\Gamma}^\alpha_{\mu\nu , \alpha}
       - \hat{\Gamma}^\alpha_{\mu\alpha , \nu} +
\hat{\Gamma}^\alpha_{\alpha\lambda}\hat{\Gamma}^\lambda_{\mu\nu} -
\hat{\Gamma}^\alpha_{\mu\lambda}\hat{\Gamma}^\lambda_{\alpha\nu}\rp\,,\label{r_def}
\ee
$\hat{\Gamma}$ being the Christoffel connection of $\hat{g}_{\mu\nu}$,
generalizes $f(R)$ theories in such a way that when $C(\R)=1$ one recovers the metric $f(R)$ theory, and when $C(\R)=f'(\R)$, the theory represents an improved variant of the Palatini-$f(\R)$ gravity\footnote{To obtain precisely the usual Palatini-$f(\R)$ theory, one may impose the constraint instead upon the connection \cite{Koivisto:2011vq}. One may consider also a metric-Palatini hybrid theory \cite{Harko:2011nh}.}. 
Furthermore, the nonminimally coupled curvature theories \cite{Koivisto:2005yk}
\be \label{nmc}
S_{nmc} = \int d^4 x \sqrt{-g} \lb f_1(R) + f_2(R)\kappa^2\mathcal{L}_m \rb\,,
\ee
can be mapped to the action (\ref{action}) written in the $C$-frame of these theories, defined by the conformal 
transformation $g_{\mu\nu} \rightarrow C(\R)g_{\mu\nu}$,
when one identifies\footnote{Stricly speaking this is assuming that the matter lagrangian is homogeneous in the metric, but it is in this 
context the nonminimally coupled theories are usually discussed in. We assumed dust-like matter above.} $f_1(R)=f(r(R))/C^2(r(R))$ and $f_2(R)=C(r(R))$ where $r(R)$ is the solution to 
the equation $r/C(r)=R$. All of the three types of $f(\R)$ theories - the metric, the Palatini and the nonminimally coupled versions 
have been very extensively studied in the literature \cite{DeFelice:2010aj}. 
The simple starting point (\ref{action}) contains them all and, between and beyond them, completely new theories.

\subsection{The PPN parameters in the limiting models}

In this paper we study the post-Newtonian parameters in the unifying framework provided by the action (\ref{action}). It is then useful to review the state of art in 
the limiting 
$f(R)$ gravities. After some initial debates \cite{Faraoni:2006hx}, it was settled that metric $f(R)$ gravity, as corresponding to the $\omega_{BD}=0$ scalar-tensor theory, 
features $\gamma=1/2$ in the massless limit that is relevant to the dark energy alternatives \cite{Chiba:2006jp}. One could perhaps however render the field locally 
massive enough by exploiting the so called chameleon mechanism in particular models \cite{Faulkner:2006ub,Henttunen:2011tr}. 

In the Palatini theories, the vacuum solutions reduce to GR and a cosmological constant, and if the value of the latter is given by the dark energy scale, one surely reproduces the 
Solar system predictions of GR to a sufficient accuracy. However, there is still some controversy about the validity of these vacuum solutions \cite{Olmo:2011uz}. From our viewpoint the 
ambiguities in the predictions of the theory can be traced down to inherent inconsistency, which hinders one from considering gradients of matter fields at 
small scales \cite{Koivisto:2005yc,Barausse:2007ys}. The action (\ref{action}), also when $C=f'$, however provides a consistent theory at all scales and thus the prescription outlined 
below will uniquely fix also the Solar system predictions for these (potentially) viable versions of the Palatini models.  

The Newtonian limit of the nonminimally coupled curvature theories remains to be clarified.
In ref. \cite{Bertolami:2008im} it was claimed that theories of the form (\ref{nmc}) share the PPN limit with GR. However this conclusion was not based on a proper derivation and 
indeed we will here find a different result. It is straightforward to show that, since (\ref{nmc}) has the biscalar representation
\be
S_{nmc} = \int d^4 {x} \sqrt{-g} \lb \psi R - 2V(\phi,\psi) + f_2(\phi)\kappa^2\mathcal{L}_m\rb\,,
\ee
where $2\kappa^2 V(\phi,\psi)=\phi\psi-f_1(\phi)$, one can write it, in the Einstein frame, as the biscalar-tensor theory
\be \label{e_nmc}
S^*_{nmc}=\sqrt{-g^*}\lb R^* - 2\sigma_{ij}\varphi^{i}_{,\mu}\varphi^{j,\mu} - 4U(\varphi^k) - f_*(\varphi^k)\kappa^2\mathcal{L}_m^*\rb\,,
\ee
where the transformations $g_{\mu\nu}^*=\psi g_{\mu\nu}$ and $\varphi^1=\sqrt{3}\log{\lp\psi\rp}/2$, $\varphi^2=\phi$ were performed. When we assume $\mathcal{L}_m$ is homogeneous of degree $n$ in the metric, we can write
\be \label{e_def}
f_*(\varphi^k) = f_2(\varphi^2)e^{-2\frac{n+2}{\sqrt{3}}\varphi^1} \equiv A^\frac{1}{4}\,.
\ee
Because the field matrix $\sigma_{ij}$ is singular,
\begin{equation} \label{sigma}
\sigma = \left(
\begin{array}{cc}
1 & 0 \\
0 & 0
\end{array} \right)\,,
\end{equation}
one cannot invert it and straightforwardly plug to model into the formalism of Damour and Esposito-Farese \cite{Damour:1992we} to obtain the post-Newtonian expansion for the 
two-field action. This is of course due to one of 
the fields being nondynamical. The way around this proposed in ref. \cite{Bertolami:2008im} was to add an antisymmetric piece $a$ to the matrix as
\begin{equation} \label{a_sigma}
\sigma = \left(
\begin{array}{cc}
1 & a \\
-a & 0
\end{array} \right)
\end{equation}
However, in general the results will depend upon the antisymmetric part when one exploits this method. This can be easily seen by adding the antisymmetric part in some other frame (in 
particular, any which is nondiagonal in the fields) and repeating the calculation. This shows that the result is not meaningful.

In fact, this can be seen by directly proceeding from (\ref{e_nmc}). Let us recall the definition of variables \cite{Damour:1992we}
\be
\alpha_i \equiv \frac{\partial \log{A}}{\partial \varphi^i}\,, \quad \alpha^2 = \sigma^{ij}\alpha_i \alpha_j\,.
\ee
Looking now at the action (\ref{e_nmc}) and the definition (\ref{e_def}), one notes that the matter is coupled nonminimally to both of the fields, and thus the effective coupling 
function $A$ should be considered to depend on both scalars (we also disagree on this point with ref.\cite{Bertolami:2008im} which set $\alpha_2=0$). We obtain that $\al_1=-(n+2)/(2\sqrt{3})$ 
and $\al_2=\lp\log{f_2(\phi)}'\rp^2/4$ and hence \be
\gamma-1=-2\frac{\alpha^2}{1+\alpha^2} = -2 \frac{\lp\log{f_2(\phi)}'\rp^2}{4a^2+\lp\log{f_2(\phi)}'\rp^2}\,.
\ee
Thus we find that the result depends upon the arbitrary parameter $a$ due to the illegitimate procedure of replacing (\ref{sigma}) with (\ref{a_sigma}).

A correct way to deal with such a theory with a nondynamical degree of freedom is to integrate it away and then follow the usual steps considering the one remaining dynamical field. We shall turn to this in the following.

\section{C-theory as scalar-tensor theory}
\label{params}

As was shown in the original paper \cite{Amendola:2010bk}, the C-theory lagrangian appearing in (\ref{action}) can be reformulated as a scalar-tensor theory involving two fields 
$\phi$ and $\xi$ as
\be \label{bst}
2\mathcal{L} = \xi R - \frac{3}{2}\xi\frac{\lp\partial C\rp^2}{C^2} + \frac{3}{C}\lp\partial_\mu C\rp\lp\partial^\mu \xi\rp - \xi\phi 
+ f(\phi) + 2\mathcal{L}_m\,. 
\ee
Here $C$ is understood as a function of $\phi$ and the matter lagrangian $\mathcal{L}_m$ is minimally coupled to gravity. The Solar 
system experiments effectively probe the vacuum metric outside a spherical source. We are looking at perturbative corrections to the Schwarzchild metric 
where sources can be neglected. Then, by studying the equations of motion ensuing from the lagrangian (\ref{bst}), one can show that there exists an 
algebraic relation between the two fields,
\be
\xi = \frac{C(\phi)f'(\phi)-2C'(\phi)f(\phi)}{C(\phi)-C'(\phi)\phi}\,.
\ee
Thus one can solve for $\phi=\phi(\xi)$ and use the result to rewrite the theory (\ref{bst}) as a single-field scalar-tensor theory
\be \label{stt}
\mathcal{L} = \frac{1}{2}\xi R - \frac{\omega(\xi)}{2\xi}\lp\partial\xi\rp^2 - V(\xi)\,, 
\ee
where the Brans-Dicke function reads, explicitly,
\be \label{bd}
\omega(\xi) = 3\xi\lb\frac{\xi}{2} \lp\frac{d\log{C(\phi(\xi))}}{d\xi}\rp^2-\lp\frac{d\log{C(\phi(\xi))}}{d\xi}\rp\rb\,.
\ee
When one considers very light fields, so that the $V(\xi)=[(\xi\phi(\xi)-f(\phi(\xi))]/(2\kappa^2)$ can be neglected, the PPN parameters can be deduced 
solely from the function $\omega(\xi)$ in the canonical form of the theory (\ref{stt}). In this massless limit, the two nontrivial parameters are
\ba 
\beta-1 & = & \frac{\omega'(\xi)}{\lp 3+2\omega(\xi)\rp\lp 4+2\omega(\xi)\rp}\Bigg\|_{\xi=\xi_0}\,, \\
\gamma-1 & = & -\frac{1}{2+\omega(\xi)}\Bigg\|_{\xi=\xi_0}\,, 
\ea
where $\xi_0$ corresponds to to present cosmological value of the scalar field. Thus we have a completely specified method to determine the weak-field corrections 
relevant at the Solar system, given any functions $f(\R)$ and $C(\R)$. 

We readily see that the well-known result $\gamma=1/2$ is obtained in the metric limit where $C$ is a constant, since then $\omega=0$. It is quite interesting that in 
general the parameter $\gamma$ becomes a function of the scalar field, which allows in principle to find nontrivial viable models. These should have 
sufficiently large $\omega(\xi_0) > 40000$ or so \cite{lrr-2006-3}. 

\subsection{Class A models}   

Let us consider the example
\be \label{pow}
f(\R) = f_0 \R^n\,, \quad C(\R) \sim \R^m\,.
\ee
The Brans-Dicke function in (\ref{bd}) becomes then
\be
\omega_{BD} = \frac{3m}{2(n-1)}\lb 2\lp n-1 \rp - m\rb\,.
\ee
The parameter $\Lambda=0$ has thus its GR value, and 
\be
\gamma-1 = -\frac{2\lp n-1\rp^2}{3m^2-6m\lp n-1\rp+4\lp n-1\rp^2}\,.
\ee
The class of models is special because it the parameters are constant regardless of the dynamics of the field. 
Remarkably, any power $m$ for the conformal factor $C$ is compatible with the Solar system constraints, as long as the dependence upon the curvature is 
nearly enough linear, $f \sim \R^{1+\epsilon}$. In particular, when the action is Einstein-Hilbert, $n=1$, GR predictions are reproduced identically. 

We expect still nontrivial cosmological modifications, since in the presence of matter the equation of motion is, when $n=1$,
\be
\lp\frac{1}{m}-1\rp\xi+\lp 2-\frac{1}{m}\rp f_0=\kappa^2\frac{\xi}{\phi}T\,.
\ee
We see that in general the scalar field can be dynamical, but in vacuum $T=0$ it reduces to a constant $\xi_{vac}=(2m-1)f_0/(m-1)$. Thus it is possible to reconcile new cosmological 
effects with the GR predictions at the Solar system level. We most probably could however distinguish these models from GR at higher orders of the PPN expansion, but that is beyond the 
scope of the present study.   

\subsection{Class B models}

Consider a variation of the latter case,
\be \label{power2}
f(\R)=\R\,, \quad C(\R) = 1+ \lp\frac{\R}{\R_0}\rp^m\,.
\ee
The PPN parameters are given explicitly in the appendix \ref{par2}. 
In this case, the corrections vanish at the point $\xi=2$,  
\be
\gamma-1=-\frac{1}{6}(\xi-2)^2-\frac{1}{3}(m-1)(\xi-2)^3 + \mathcal{O}\lp(\xi-2)^4\rp\,,
\ee
\be
\beta-1 = -\frac{1}{12}(\xi-2)+\frac{1}{4}(\xi-2)^2+   + \mathcal{O}\lp(\xi-2)^3\rp\,.
\ee
The point $\xi\rightarrow 2$ corresponds to the limit of infinite $\R\rightarrow \infty$. It should be noted that this
quantity is different from the metric spacetime curvature $R$, which should be very small in the appropriate
limit. However, it is not clear such a configuration may be naturally arranged.

We find similar result by looking at the exponential form 
\be \label{exp}
f(\R) = \R\,, \quad C(\R) \sim \exp{\lp k\R\rp}\,.
\ee
The effective Brans-Dicke parameter is then in vacuum
\be
\omega=\frac{3\xi\lp 8 + (2\xi-7)\xi\rp}{2(\xi-2)^2}\,.
\ee
Again, the corrections vanish at the point $\xi=2$,  
\be
\gamma-1=-\frac{1}{6}(\xi-2)^4-\frac{1}{6}(\xi-2)^5 + \mathcal{O}\lp(\xi-2)^6\rp\,,
\ee
\be
\beta-1 = -\frac{1}{6}(\xi-2)^3+\frac{5}{24}(\xi-2)^4 + \mathcal{O}\lp(\xi-2)^5\rp\,.
\ee
The complete parameters are given in appendix \ref{par3}.

One more example which yields qualitatively very similar results is given by the example
\be \label{exp2}
f(\R) = \R\,, \quad C(\R) \sim \exp{\lp 1+\lp\R/\R_0\rp^m\rp}\,.
\ee
What one finds again in this case is that the post-Newtonian corrections vanish at $\xi=2$, which however implies an infinite $\R$. 

\subsection{Class C models}

Finally, we consider the parameterization
\be \label{class_c}
f(\R) = \R + \al \lp \frac{\R}{\R_0}\rp^n \R\,, \quad \log{C(\R)} = 1 + \lp \frac{\R}{\R_0}\rp^n\,.
\ee
We chose the same exponent $n$ to simplify analytic calculations. The purpose of this example is to illustrate a third possible kind of
qualitatively different behaviour of the PPN parameters. To wit, the parameters need not be constant nor $\R$ large, while there still exists solutions 
which reproduce precisely GR predictions up to the leading post-Newtonian order.
From the form of the Brans-Dicke function (\ref{bd_c}) one sees that the corrections vanish at the point
\be
\xi=\frac{1}{n}\lb\lp3-n\rp\al+2n\pm2\sqrt{2\alpha}\sqrt{\alpha+(1-\al)n}\rb\,.
\ee
This corresponds to the value of the scalar $\R$
\be
\R = \lb \frac{2\al \pm \sqrt{2\al}\sqrt{\al+(1-\al)n}}{2\al n}\rb^\frac{1}{n}\R_0\,,
\ee
that is obviously nonsingular.

\section{Conclusions}

We provided a recipe to compute the PPN parameters for C-theory. According to the behaviour of these parameters, we classified C-theories into three classes. 
In class A models, the parameters are constant. The metric $f(R)$ models belong to this class, characterized by $\ga=1/2$, but there are also viable models. In particular, the 
power-law models (\ref{pow}) can mimic GR when $n$ is close to unity. In class B models, the GR limit
seems more difficult to produce as it does not correspond to finite solution for the scalar curvature. In class C models, the PPN parameters are functions of the scalar field in such a 
way that the GR limit can be obtained when both of the two effective scalar fields are finite. 

These results provide further motivation to explore the cosmological implications of the new  theories included within the unifying framework of C-theories (on implications of nonmetricity to chaotic inflation, see Ref. \cite{Enqvist:2011qm}). 
It also remains to be seen whether those theories can be observationally distinguished by testing their predictions at the strong
field regime or at the higher orders of the PPN expansion.

\acknowledgments

The author would like to thank Luca Amendola and Kari Enqvist for useful discussions, and the anonymous referee and Marit Sandstad for 
helpful comments on the manuscript. 

\begin{widetext}

\appendix

\section{Specific models}
\label{app}

\subsection{Power-law model}
\label{par2}

For the model (\ref{power2}) the parameters have the following form.
\be 
\gamma-1 = -\frac{2 (\xi-2)^2}{\xi \left(3 m^2 \xi (\xi-2)^2-6 m ((\xi-2) \xi+2) (\xi-2)+\xi (3 (\xi-4) \xi+25)-28\right)+16}\,.
\ee
\be
\beta-1=
\frac{(\xi-2) \left(m (\xi-2)^2-(\xi-4) \xi-2\right)}{((m-1) (\xi-2) \xi-2) \left(\xi \left(3 m^2 \xi (\xi-2)^2-6 m ((\xi-2) \xi+2) (\xi-2)+\xi (3 (\xi-4)
   \xi+25)-28\right)+16\right)}\,.
\ee

\subsection{Exponential model}
\label{par3}

For the model (\ref{exp}) the parameters have the following form.
\be
\gamma-1 = -\frac{2 (\xi-2)^4}{\xi \left(\xi \left(4 \xi^2-26 \xi+75\right)-104\right)+64}\,.
\ee
\be
\beta-1 = -\frac{(\xi-2)^3 (\xi+2)}{((\xi-3) \xi+4) \left(\xi \left(\xi \left(4 \xi^2-26 \xi+75\right)-104\right)+64\right)}\,.
\ee

\subsection{The double power-law model}

The Brans-Dicke function computed for the model (\ref{class_c}) is
\be \label{bd_c}
\omega(\xi) = \frac{3 \xi \left(n \pm \frac{n (\al (n-3)+n (\xi-2))}{\sqrt{(\al (n+1)+n (\xi-2))^2-8 \al n (\xi-1)}}\right) \left(\xi \left(n \pm \frac{n (\al (n-3)+n 
(\xi-2))}{\sqrt{(\al 
(n+1)+n(\xi-2))^2-8 \al n (\xi-1)}}\right)-8 \al n\right)}{32 \al^2 n^2}\,.
\ee

\end{widetext}

\bibliography{cppn}

\end{document}